\pdfoutput=1
\documentclass[aps,prl,twocolumn,
		superscriptaddress,a4paper,
   floatfix]{revtex4}
\usepackage[latin1]{inputenc}
\usepackage{graphicx}
\usepackage{amsmath,amssymb}
\usepackage{datetime}
\usepackage{pgfpages}
\usepackage{tikz}

\begin{document}

\title{Observing the Drop of Resistance in the Flow of a Superfluid Fermi Gas}

\author{David Stadler}
\author{Sebastian Krinner}
\author{Jakob Meineke}
\author{Jean-Philippe Brantut}
\author{Tilman Esslinger}
\affiliation{Institute for Quantum Electronics,  ETH Zurich, 8093 Zurich, Switzerland}

\maketitle

{\bf The ability of particles to flow with very low resistance is a distinctive character of a superfluid or superconducting state and led to its discovery in the last century \cite{leggett_quantum_2006, delft_discovery_2010}. While the particle flow in liquid Helium or superconducting materials is essential to identify superfluidity or superconductivity, an analogous measurement has not been performed with superfluids based on ultracold Fermi gases. Here we report on the direct measurement of the conduction properties of strongly interacting fermions, and the observation of the celebrated drop of resistance associated with the onset of superfluidity. We observe variations of the atomic current over several orders of magnitude by varying the depth of the trapping potential in a narrow channel, which connects two atomic reservoirs. We relate the intrinsic conduction properties to thermodynamic functions in a model-independent way, making use of high-resolution in-situ imaging in combination with current measurements. Our results show that, similar to solid-state systems, current and resistance measurements in quantum gases are a sensitive probe to explore many-body physics. The presented method is closely analogous to the operation of a solid-state field-effect transistor. It can be applied as a probe for optical lattices and disordered systems, and paves the way towards the modeling of complex superconducting devices.}

Over the last decade, cold atoms have emerged as a many-body system with a uniquely high level of control \cite{bloch_many-body_2008}. Experiments have shown that interacting atomic Fermi gases, analogous to electrons in a solid, can display superfluidity \cite{giorgini_theory_2008}. The equilibrium properties of those gases have been measured with high precision \cite{horikoshi_measurement_2010, nascimbene_exploring_2010, ku_revealing_2012} and the superfluid character of the ground state has been investigated via the response to external perturbations \cite{miller_critical_2007} and rotation \cite{zwierlein_vortices_2005}, in the same way as for Bose-Einstein condensates \cite{madison_vortex_2000, matthews_vortices_1999, raman_evidence_1999, burger_superfluid_2001, amo_superfluidity_2009}. With the new techniques to create and observe directed currents \cite{ramanathan_superflow_2011,brantut_conduction_2012} it is now possible to study transport properties of mesoscopic systems directly analogous to electronic devices \cite{seaman_atomtronics:_2007}.

In this work, we investigate the conduction properties of strongly interacting fermions flowing through a quasi two-dimensional, multimode channel, which connects two atomic reservoirs \cite{brantut_conduction_2012}. As illustrated in figure 1, the atomic current in the channel is controlled using a repulsive potential created by an off-resonant laser beam. In analogy with an electronic field-effect transistor, this gate potential controls the chemical potential in the channel while keeping the temperature imposed by the reservoirs unchanged. With the gate potential as a control parameter, we measure the current through the channel over a large dynamic range and determine the density distribution in the channel region. This allows us to observe the onset of frictionless flow of strongly interacting fermions. These measurements are compared to the case of a weakly interacting Fermi gas.

\begin{figure}
	\centering
		\includegraphics[width=8.9cm]{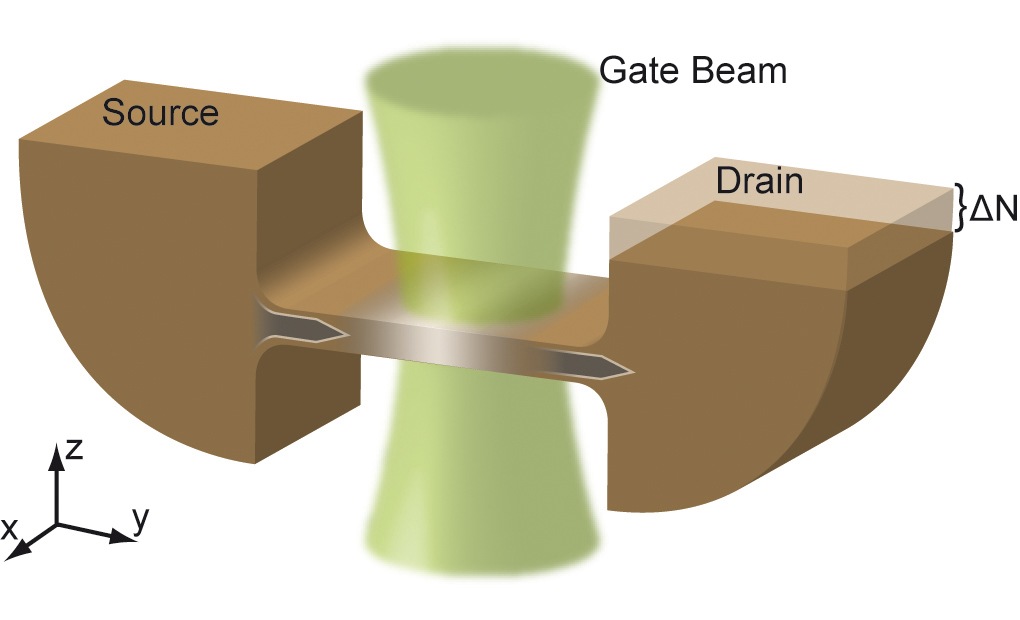}
	\caption{Principle of the experiment. Two atomic reservoirs (source and drain) are connected by a quasi-two dimensional conducting channel. The curved shape of the reservoirs indicates the harmonic confinement along the $y$-axis. An atom number imbalance $\Delta N$ between source and drain drives an atom current through the channel, indicated by the arrows. A repulsive laser beam (gate beam) propagating along the $z$-axis is focused on the channel. It creates a repulsive potential with a gaussian envelope and a tunable amplitude. The lighter region in the channel indicates the reduced density due to the repulsive potential. }
	\label{fig:setup}
\end{figure}

Our experiments are performed with strongly and weakly interacting quantum degenerate gases of fermionic $^6\mathrm{Li}$ atoms, equally populating the lowest two hyperfine states. To obtain a strongly interacting gas, the atoms are placed in a homogeneous magnetic field of 834\,G where interactions are attractive and lead to the formation of pairs, while a weakly interacting gas is studied at a field of 475\,G. The atoms are radially confined in the $x$-$z$ plane by an optical dipole trap oriented along the $y$-axis with a $1/e^2$ beam radius of $22(1)$\,$\mu$m. Along the $y$-direction, the curvature of the magnetic field yields a harmonic confinement with a frequency of $\omega_y= 2 \pi \cdot 32(1)\,\mathrm{Hz}$. To engineer the reservoirs, we split the cloud into two parts using a repulsive laser beam at a wavelength of $532 \, \mathrm{nm}$ that points along the $x$-direction (beam not shown in figure 1). The intensity profile of this beam has a holographically imprinted nodal line along the $y$-axis. As a result, a channel in the $x$-$y$ plane is formed, which confines the atoms along the $z$-direction with a center trap frequency of $2.9$\,kHz.  The gate potential is created by another laser beam at $532 \, \mathrm{nm}$ that is sent along the $z$-axis onto the channel and has a waist of $18 \, \mu \mathrm{m}$. We refer to the maximum of the repulsive potential created by this beam as the gate potential $U$. Along the $z$-axis, a high-resolution microscope objective is used for in-situ absorption imaging of the atoms in the channel. The atom number in the reservoirs is measured by absorption imaging along the $x$-direction. By creating an atom number imbalance between the two reservoirs, we create a chemical potential bias that induces a current through the channel \cite{brantut_conduction_2012}.
 
Figure 2A presents an example of the time evolution of the relative number imbalance between the two reservoirs, measured for strongly interacting (red) and weakly interacting fermions (blue), using the same gate potential of $U=525(50) \, \mathrm{nK}$. For the strongly interacting gas, an exponential fit yields a decay time of $0.076(7)\,\mathrm{s}$, which is one order of magnitude faster than the decay time of $0.70(6)\,\mathrm{s}$ obtained for the weakly interacting gas.

The reservoirs can be considered to be in quasi-thermal equilibrium during the entire decay, provided this process is sufficiently slow compared to the thermalization dynamics within the reservoirs. Thus, except for the lowest gate potentials, we interpret the exponential decay of the imbalance as a resistance measurement analogous to the discharge of a capacitor, where the resistance of the channel is proportional to the decay time $\tau$. The proportionality factor is determined by the reservoirs and remains constant as the gate potential is varied \cite{brantut_conduction_2012}. The dimensionless resistance $r = \tau \cdot \omega_y$ is shown in figure 2B as a function of the gate potential $U$. Here, $\omega_y$ is the frequency of the harmonic confinement along the $y$-axis, which provides a natural time scale for the motion of atoms along this direction. For decreasing gate potential the weakly interacting Fermi gas (blue) shows a decrease of resistance reaching a minimum value of $r \approx 35$ for zero gate potential. For high gate potentials the resistance for both interaction strengths are comparable, yet the strongly interacting gas (red) shows a much faster drop of resistance below $0.7\, \mu\mathrm{K}$. At a gate potential of $0.23(2) \, \mu\mathrm{K}$ the resistance differs by a factor of about $25$ from the weakly interacting gas. As $r$ approaches unity (below $0.23 \, \mu\mathrm{K}$) the decay time $\tau$ becomes equal to the time scale of the internal dynamics of the reservoirs, set by the trap frequency along the $y$-direction. In this regime, we cannot interpret our strongly interacting data sets in terms of a resistance measurement because the reservoirs do not remain in thermal equilibrium at each point in time, i.e. the resistance has dropped below our measurement capabilities. This gives rise to deviations from the exponential decay.

\begin{figure}
	\centering
		\includegraphics[width=8.9cm]{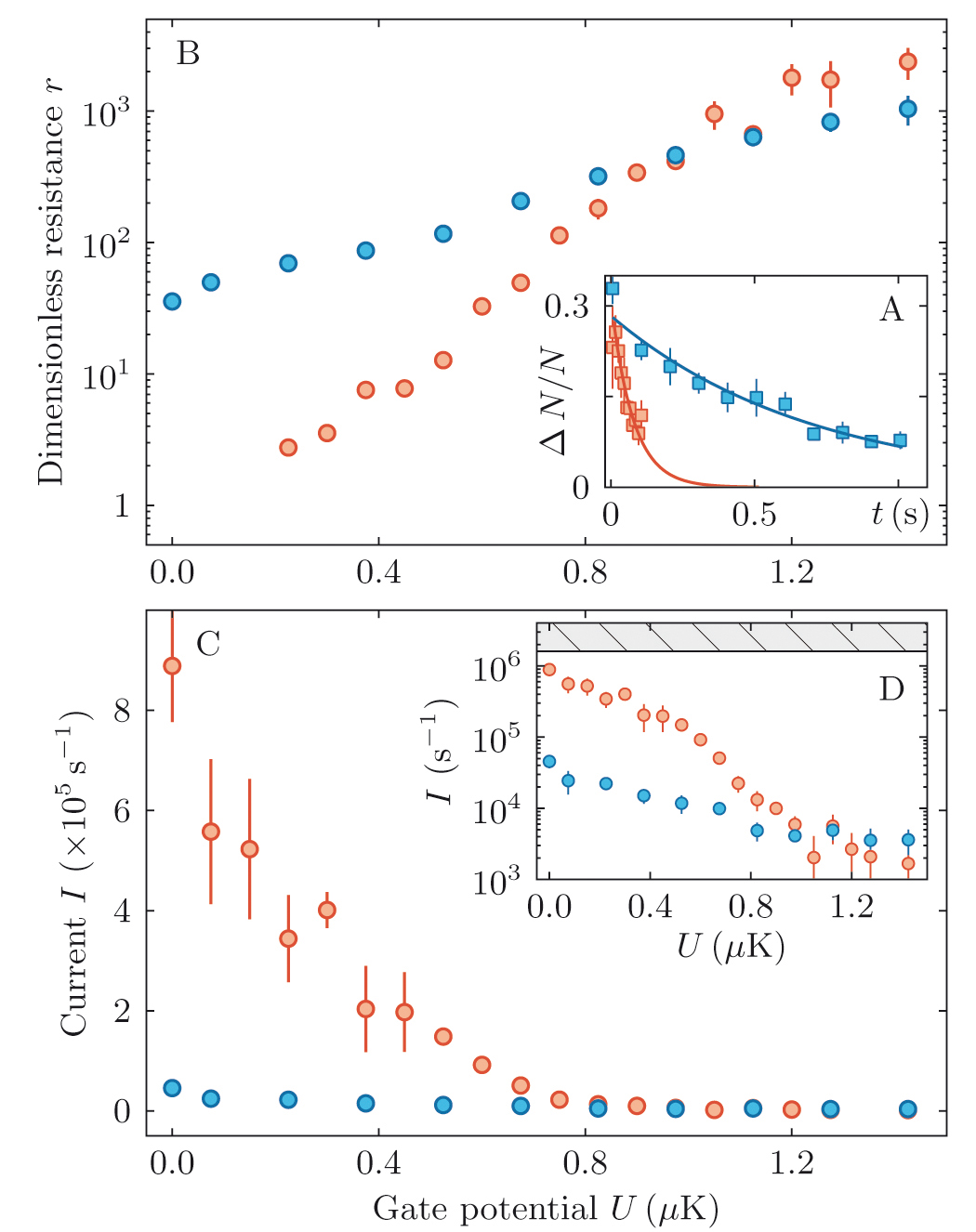}
	\caption{Conduction properties through the channel. Red and blue data points correspond to the strongly and weakly interacting gas, respectively. A: Decay of the relative atom number imbalance between source and drain as a function of time with a gate potential $U = 525(50) \,\mathrm{nK}$. The solid lines are exponential fits without offset. B: Dimensionless resistance $r$ as a function of gate potential. The data points that are shown are those for which the decay is exponential. C: Atom current as a function of the gate potential $U$. A large increase of the current appears for the strongly interacting gas below $U \approx 0.7 \, \mu \mathrm{K}$. D: Atom current in logarithmic scale. The dashed region indicates the maximum current allowed by the internal dynamics of the reservoirs (see Text). The error bars show the statistical errors.}
	\label{fig:current}
\end{figure}

In addition to the resistance, we also estimate the current through the channel using a linear fit to the initial part of the decay (see Methods). This measurement does not rely on the thermalization of the reservoirs and thus can also be applied to cases where the reservoirs are not fully in quasi-thermal equilibrium. Figure 2C shows the current $I$ as a function of the gate potential for the strongly interacting gas (red) and the weakly interacting gas (blue). Contrary to the weakly interacting gas, the strongly interacting gas shows a fast increase of the current below $0.7\, \mu\mathrm{K}$. For the lowest gate potentials the current is limited by the conservation of energy. The limit is reached when the potential energy introduced by the initial imbalance is full converted into kinetic energy, as for example in undamped dipole oscillations. It is represented by the shaded region in figure 2D, where we show the current in logarithmic scale. Remarkably, the observed current is very close to that limit, meaning that the strongly interacting Fermi gas flows as if there was no constriction or gate potential at all. This is the expected behavior of a superfluid.

Whilst the current depends on the atomic density in the channel, the transport properties are characterized in a density independent way by the drift velocity. To extract this quantity, we first use high-resolution in-situ imaging to measure the atomic line-density in the channel. The measured line-density as a function of the gate potential is shown in figure 3B. As expected from its higher compressibility \cite{ku_revealing_2012,bartenstein_crossover_2004}, the strongly interacting gas reaches larger line-densities. For each value of the gate potential, we then divide the measured current by the corresponding line-density yielding the drift velocity.

\begin{figure}
	\centering	
\includegraphics[width=8.9cm]{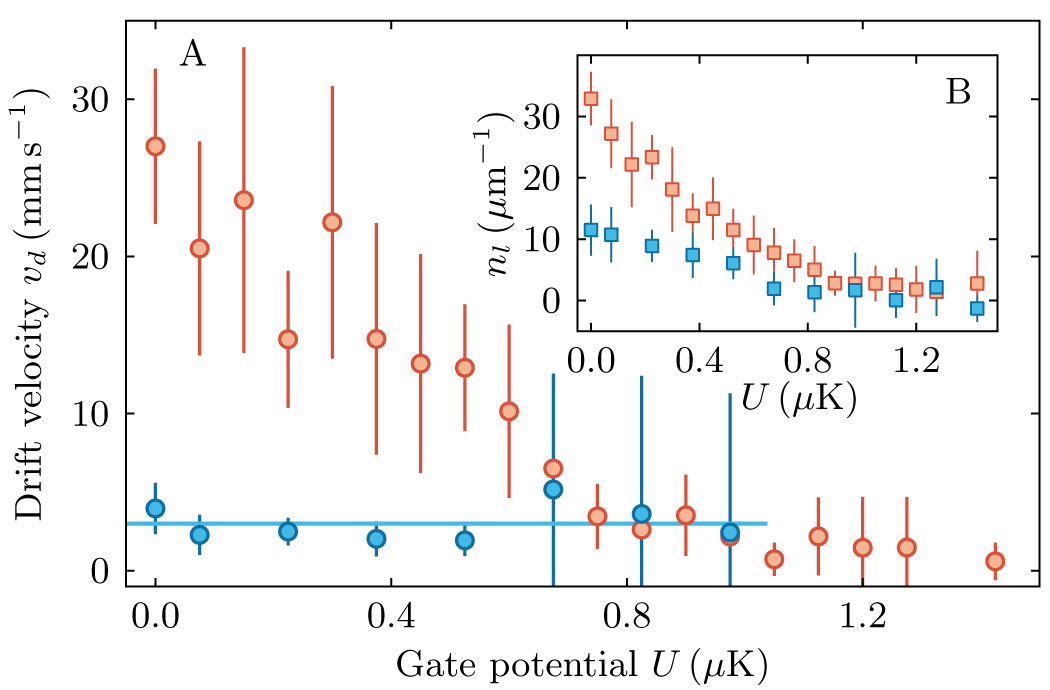}
	\caption{Density independent conduction properties through the channel. Red and blue data points correspond to strongly interacting and weakly interacting atoms, respectively. A: Drift velocity as a function of gate potential. The blue line is a guide to the eye. The points corresponding to the three highest values of the gate potential are omitted in the weakly interacting case since the density is almost zero. B: Line-density measured in-situ in the channel as a function of gate potential (see Methods). The error bars represent statistical errors.}
	\label{fig:lineDens_and_vd}
\end{figure}

The drift velocities as a function of gate potential are presented in figure 3A. The weakly interacting gas shows a constant drift velocity, as expected for a normal conductor. In contrast, the drift velocity for the strongly interacting gas increases significantly below $0.7\, \mu \mathrm{K}$. This demonstrates that the large increase of the current, as seen in figure 2, is not simply caused by the higher density of the strongly interacting gas in the channel. This reveals a change in the nature of the transport process, which is expected at the onset of superfluidity.

We now relate the conduction properties to a thermodynamic parameter by replacing the gate potential scale, which is specific for our system, by the thermodynamic potential. To this end, we use the high-resolution images of the gas in the channel, which give us access to the equation of state \cite{nascimbene_exploring_2010,ku_revealing_2012,horikoshi_measurement_2010}. The gas in the channel is in the crossover regime between two and three dimensions, where the equation of state naturally relates the column density $n_{\mathrm{col}}$ to the chemical potential \cite{orel_density_2011,dyke_crossover_2011} (see Methods). From the in-situ absorption images of the channel for different gate potentials we get $n_{\mathrm{col}}(U)$ at fixed temperature, which is imposed by the reservoirs. Integrating this relation over the known variations of the gate potential yields the thermodynamic potential $\Omega(U) = \int_U n_{\mathrm{col}}(V) dV$. It would be equal to the pressure in a purely two-dimensional gas. We normalize $\Omega$ by the pressure of a 2-dimensional ideal Fermi gas at zero temperature $\Omega_0 = \pi\hbar^2 n_{\mathrm{col}}^2/m$ and obtain a model-independent thermodynamic scale, analogous to the three-dimensional situation discussed in \cite{ku_revealing_2012}. This allows us to convert the gate potential into a thermodynamic quantity, even though the gas in the channel is not expected to be in the universal regime \cite{ho_universal_2004} but rather in the confinement-dominated regime \cite{petrov_interatomic_2001}, where most of the thermometry techniques cannot be applied directly \cite{ku_revealing_2012,luo_measurement_2007}.

\begin{figure}
	\centering
		\includegraphics[width=8.9cm]{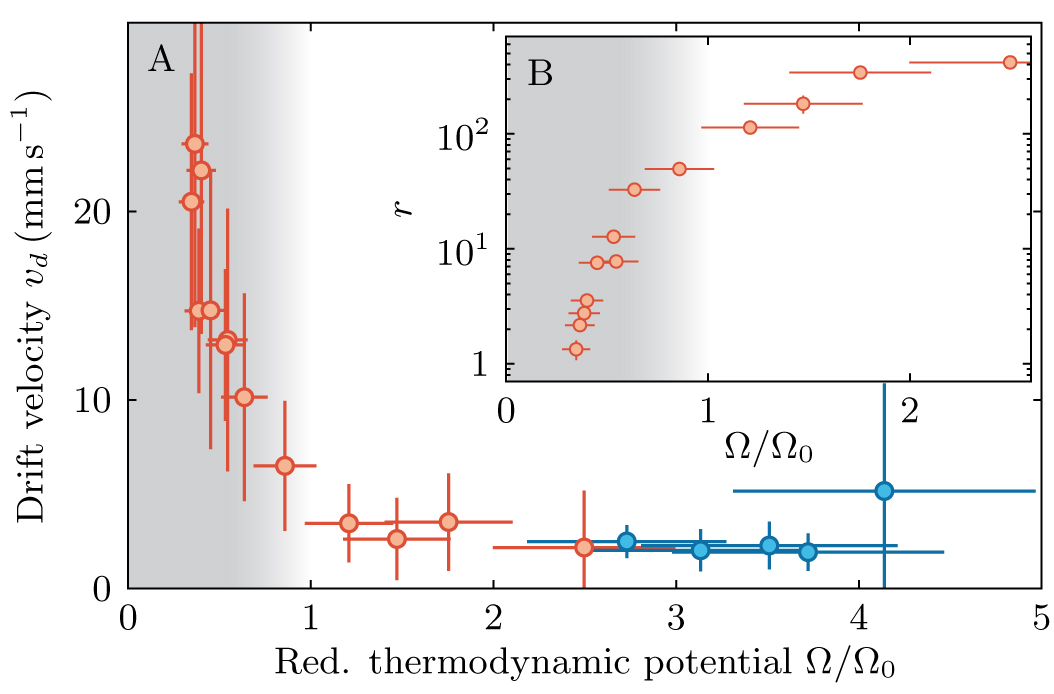}
	\caption{Conduction properties as a function of thermodynamic potential. A: Drift velocity as a function of the reduced thermodynamic potential $\Omega / \Omega_0$ for the strongly interacting (red) and weakly interacting (blue) Fermi gas. The gray shaded region indicates the regime where the gas is superfluid. B: Dimensionless resistance as a function of $\Omega / \Omega_0$ in logarithmic scale for the strongly interacting gas, showing the drop of resistance. Error bars represent statistical errors.}
	\label{fig:reducedTherm}
\end{figure}

The drift velocity as a function of reduced thermodynamic potential is shown in Figure 4A. The strongly interacting gas (red) shows a pronounced increase of drift velocity below $\Omega / \Omega_0 \approx 1$, indicating the onset of superfluidity. This illustrates the high sensitivity of transport measurements to many-body effects in strongly correlated quantum gases. Figure 4B presents the resistance as a function of $\Omega / \Omega_0$ for the strongly interacting Fermi gas. Below $\Omega / \Omega_0 \approx 1$ the resistance shows a clear kink even in logarithmic scale. This is the counterpart of the drop of resistance observed in superconductors.

The geometry of our experiment is reminiscent of weak links in superconductors \cite{likharev_superconducting_1979}. It operates in a regime where the channel is ballistic in the normal case, but where the link is much longer than the healing length of the superfluid, complementary to the Josephson regime explored with Bose-Einstein condensates \cite{albiez_direct_2005,leblanc_dynamics_2011}.

Our setup allows the investigation of superfluidity and supercurrents in a variety of configurations by projecting a designed potential through the microscope onto the channel \cite{zimmermann_high-resolution_2011}. This opens the way towards the cold-atom modeling of complex, superconducting devices.

\section*{Methods Summary}

A balanced mixture of the two lowest hyperfine states of $^6$Li ($\approx 8\cdot 10^4$ atoms) is prepared by all-optical evaporation. Final temperatures are $\approx 0.1\, T_F$ (strongly interacting gas) and $\approx 0.3 \, T_F$ (weakly interacting gas). For the strongly interacting gas the evaporation is performed at a magnetic field of 795\,G (scattering length $3500 \, a_0$, $a_0$ is the Bohr radius), then the field is adiabatically ramped to 834\,G, at the s-wave Feshbach resonance. The weakly interacting gas is cooled at 300\,G, then the field is ramped to 475\,G (scattering length $-100 \, a_0$). The trap frequency along the $y$-axis is $\omega_y = 2\pi\cdot 32(1)$ and $\omega_y = 2\pi\cdot 25(1)$ for the strongly and weakly interacting gas, repectively. To induce an atom current, we create a number imbalance between the two reservoirs by shifting the trapping potential along the $y$-direction with a magnetic field gradient of $0.25\,\mathrm{G} \cdot \mathrm{cm}^{-1}$. After switching off the gradient within $10\,\mathrm{ms}$, we monitor the decay of the number imbalance. The number imbalance and the total atom number are obtained from absorption images along the $x$-axis. For all data, we fit a line to the first 5 points of a measured decay curve of the relative number imbalance. We define the current as the fitted slope times half the total number of atoms in both reservoir at equilibrium. To measure the column density $n_{\mathrm{col}}$, as well as the line-density at the center of the channel for different gate potentials we take in-situ absorption images of the channel through the high-resolution microscope in the absence of current. We apply light pulses of $5$\,$\mu$s and a saturation of $\simeq0.1$. Using local density approximation gives the equation of state $n_{\mathrm{col}}(\mu_0 - V)$ \cite{nascimbene_exploring_2010}, where $\mu_0$ is the chemical potential imposed by the reservoirs and $V$ the local gate potential. Integrating this equation with respect to the gate potential leads to the thermodynamic potential.

\section*{Methods}

\paragraph{Cloud preparation}

A quantum degenerate Fermi gas is prepared by all-optical evaporation of a balanced mixture of the two lowest hyperfine states of $^6$Li. Evaporation is performed at a magnetic field of 795\,G (where the scattering length is $3500 \, a_0$, $a_0$ is the Bohr radius) down to a trap depth of 880\,nK. This produces a Bose-Einstein Condensate of molecules. Then the trap depth is increased to 2.6\,$\mu$K in order to stop the evaporation, and the magnetic field is adiabatically ramped up to $834$\,G, where the broad s-wave Feshbach resonance is positioned. The curvature of this Feshbach field sets the trap frequency along the $y$-axis $\omega_y =2\pi\cdot 32(1) \, \mathrm{Hz}$. We obtain a strongly interacting Fermi gas of about $8\cdot 10^4$ atoms with a temperature $T \approx 0.1\, T_F$ \cite{bartenstein_crossover_2004}, where $T_F$ is the Fermi temperature. We determine the chemical potential ($\mu \approx 0.8\, \mu\mathrm{K}$) of the strongly interacting gas by measuring the size of the cloud in the trap. The weakly interacting Fermi gas is prepared using the same evaporation ramp at a magnetic field of $300$\,G. The magnetic field is then ramped up adiabatically to $475$\,G ($\omega_y =2\pi\cdot 25(1)\,\mathrm{Hz}$) where the scattering length is -100\,$a_0$. This yields atom numbers of about $6.5\cdot 10^4$ at $T \approx 0.3\, T_F$. We keep the scattering length at a small but finite value to ensure that the reservoirs remain at equilibrium during the measurement.

\paragraph{Current generation and measurement}

During evaporative cooling we create a number imbalance between the two reservoirs by having the trapping potential shifted along the $y$-direction, away from the center position of the channel. The shift is created using a magnetic field gradient of $0.25\,\mathrm{G} \cdot \mathrm{cm}^{-1}$ along the $y$-axis. Restoring the symmetry of the potential in $10\, \mathrm{ms}$ creates an atom number imbalance in the symmetric trapping configuration. This leads to a potential imbalance, inducing the atom current. To infer the atom number imbalance, as well as the total atom number, the number of atoms in each reservoir is measured by absorption imaging along the $x$-axis. This is done for variable time delays. Each measurement is repeated 3 times and averaged to reduce the noise. For all data, we fit a line to the first 5 points of a measured decay curve of the relative number imbalance. We define the current $I$ as the fitted slope multiplied with half the total atom number in both reservoirs at equilibrium. For the case where the decay is exponential we checked that fitting a line and an exponential gives the same current within the error bars.

\paragraph{Equilibrium density of the gas}

In the absence of current, we take in-situ absorption images of the cloud through the high-resolution microscope. We use light pulses of $5$\,$\mu$s with an intensity of $\simeq0.1$ of the saturation intensity. We extract the line-density of the cloud by counting the total number of atoms in a region of $18$\,$\mu$m along the $y$-axis at the center of the channel, over which the trap frequency along the $z$-axis varies by less than $10$\,\%. The variations of column density along the $x$-axis are measured by counting the atom number in patches of length $18$\,$\mu$m in the $y$-direction, and $2.4$\,$\mu$m in the $x$-direction. From the known waist of the dipole trap ($22(1)\,\mu\mathrm{m}$), we infer that the change of chemical potential within one of those patches is lower than $13.7$\,\%. All in-situ pictures are averaged $20$ times to reduce the noise. In addition, the gate beam profile is directly imaged through the same optical system, yielding a map of the gate potential.

\paragraph{Thermodynamic potential}

For each power setting of the gate beam, the in-situ column density along the $x$-axis is processed in seven patches to yield a set of curves $n_{\mathrm{col}}(V)$, where $V$ is the local gate potential in the corresponding patch. In the local density approximation, these curves belong to the same equation of state (as the confinement along the $z$-axis is the same in all patches). The curves are combined using the hypothesis that regions having the same column density have the same chemical potential, giving the equation of state $n_{\mathrm{col}}(\mu_0 - V)$ \cite{nascimbene_exploring_2010}. Here $\mu_0$ is the unknown chemical potential imposed by the reservoirs. Integrating this equation of state from $V$ to the largest gate potential (for which density is zero) gives the thermodynamic potential as a function of $U$ for a fixed (but unknown) temperature. By normalizing the thermodynamic potential to that of an ideal two-dimensional Fermi gas with the same column density, we obtain the thermodynamic scale which is used for figure \ref{fig:reducedTherm}. 

\paragraph{Confinement dominated regime in the channel}

The size of the superfluid pairs in three dimensions on the Feshbach resonance is $2.6/k_F \approx 0.6$\,$\mu\mathrm{m}$ \cite{schunck_determination_2008}, which is of the order of $\sqrt{\hbar/(m \omega_z)} \approx 0.8$\,$ \mu\mathrm{m}$, i.e. the size of the ground state of the harmonic oscillator. Considering this criterion, the superfluid is in the crossover from two to three dimensions, eventhough the chemical potential of the gas is larger than the oscillation frequency in the channel.

\bibliographystyle{naturemag}

\bibliography{paper}

\paragraph{Acknowledgments} We acknowledge enlightening discussions with Wilhelm Zwerger, Antoine Georges, Corinna Kollath and Charles Grenier. We thank Leticia Tarruell, Tobias Donner and Henning Moritz for their careful reading of the manuscript. We acknowledge Financing from NCCR MaNEP and QSIT, ERC project SQMS, FP7 project NAME-QUAM and ETHZ. JPB acknowledges support from EU through Marie Curie Fellowship. 

\paragraph{Author Contributions} All authors contributed equally to this work.

\paragraph{Author Information} The authors declare no competing financial interests. Correspondence should be addressed to Jean-Philippe Brantut (brantutj@phys.ethz.ch) and Tilman Esslinger (esslinger@phys.ethz.ch).

\end{document}